# Dimensional Reduction in Complex Living Systems: Where, Why, and How

Jean-Pierre Eckmann[1] and Tsvi Tlusty[2,3,*]

[1] Département de Physique Théorique and Section de Mathématiques, Université de Genève, CH-1211, Geneva 4, Switzerland.
[2] Center for Soft and Living Matter, Institute for Basic Science, Ulsan 44919, Republic of Korea.
[3] Departments of Physics and Chemistry, UNIST, Ulsan 44919, Republic of Korea
* tsvitlusty@gmail.com

The unprecedented prowess of measurement techniques provides a detailed, multi-scale look into the depths of living systems. Understanding these avalanches of high-dimensional data—by distilling underlying principles and mechanisms—necessitates dimensional reduction. We propose that living systems achieve exquisite dimensional reduction, originating from their capacity to learn, through evolution and phenotypic plasticity, the relevant aspects of a non-random, smooth physical reality. We explain how geometric insights by mathematicians allow one to identify these genuine hallmarks of life and distinguish them from universal properties of generic data sets. We illustrate these principles in a concrete example of protein evolution, suggesting a simple general recipe that can be applied to understand other biological systems.

## I. Introduction: the problem of dimension in biology

What is the inherent complexity of living things? One can answer this ill-defined question by specifying a **dimension** — roughly speaking, the number of independent parameters that describe a system. But that is not so straightforward. Is the ant, walking in our kitchen, a $10^9$-dimensional entity described by a billion-letter word, its genome? Or is rather the 2-dimensional trajectory that the ant draws on the kitchen's floor a more relevant depiction? [1]

---

[1] The length of ant genomes varies between 0.2-0.5 billion bp. Also note that the space of all possible 2D *trajectories* is still high-dimensional.

In our times, this problem of dimension has become a matter of vital practical importance. Driven by technological leaps, the accelerated progress of biology brought about novel tools of measurement and, with them, a flood of data in huge-dimensional spaces. Analyzing such data to understand their meaning is a grand challenge.

First, for us humans, the process of **understanding** naturally involves dimensional reduction. An essential hallmark of any good model is the compression of reality to its more "relevant" aspects. Consider an enzyme, an example we will revisit throughout this short review. One may look up the sequence of this enzyme or its high-resolution structure and plot the position of each atom and the electronic density. These high-dimensional data—proteins are composed of thousands of atoms—may be useful for detailed simulations, for example, of a substrate docking onto the enzyme's active site. On the other hand, in understanding a biochemical experiment, the relevant quantities are the kinetic parameters: two numbers, catalytic rate and affinity, suffice to describe the dynamics of many enzymatic reactions.

Second, and no less important, it is a matter of practicing science: high-dimensional spaces are extremely hard to manipulate and sample as they explode with the dimension. Thus, the huge "universe" of all possible protein sequences, whose number increases exponentially with the gene length $D$ as $4^D$, has certainly not been thoroughly explored even by 3.5 billion years of evolution[1,2]. Similarly, the combinatorial space of developmental programs explodes super-exponentially with the size of the underlying gene network[3]. The complexity of biological systems only worsens the problem. Living systems are multilayered networks operating at various space and time scales[4-6]. The basic problem is how to select the relevant layer and how to treat its coupling to other layers. No general recipe is available, and finding the relevant degrees of freedom is always context-dependent, and should reflect the specific question one tries to address.

Our experience shows that quite drastic dimensional reduction occurs in numerous living systems. Recent examples include the swimming motion of the ciliate *Tetrahymena*[7], the behavior of worms[8] and fruit flies [9], the morphology of the beaks of Darwin's finches[10,11], the spaces of plant traits[12], the branching patterns of flowers[13], gene expression in *Escherichia coli*[11,14], the biochemical phenotype space of the Rubisco enzyme[15], population dynamics of closed microbial communities[16], and allosteric coupling in proteins[17].

In this essay, we will discuss a few basic questions about this phenomenon: whether dimensional reduction is an inherent hallmark of biology, and, if so, **where** it occurs and **why**, and finally, we will touch on the practical question of **how** one may try to disentangle high-dimensional complexity—as much as one is allowed to generalize and conjecture in biology.

Specifically, we will discuss whether this reduction is in the nature of any generic large data set or to what extent it provides additional insight into the inner workings of biological systems. In a nutshell, we argue that both alternatives play a role:

- A natural mathematical simplification applies to any high-dimensional data set.

- An additional, *significant* simplification stems from the special nature of biological systems.

We illustrate this in the case of the protein, which we studied in some detail [18-21]. We will see that while dimensional reduction is omnipresent in nature and society—in chemical reaction networks as well as in stock-market data—not all reductions are the same. Some dimensional reductions seem obvious from looking with fuzzy glasses (that is, by coarse-graining), while other reductions have deeper origins, particularly learning and smoothness.

**Smoothness** captures the relative simplicity of an underlying intrinsic structure. A feature is smooth if it does not vary too abruptly in time and space, and therefore can be captured by a small number of parameters. Thus, the description of smooth systems can be compressed into low-dimensional spaces[22]. For example, the volume of a fast-growing bacterium is well-approximated by an exponential curve determined by one parameter, the growth rate. Likewise, the shapes of many organisms can be described by relatively simple mathematical functions determined by a few parameters [23]. In experiments, smoothness is often achieved by averaging over many, hopefully comparable, samples, thereby filtering high-dimensional variance, which we sometimes call "noise."

**Learning** is the process by which we, and other biological systems, represent reality by identifying and retaining only its essential features. Such reduction of irrelevant complexity marks the process of understanding. Great theories capture reality into simple forms by immense **compression**, while a theory that remains as complex as reality is impractical, as the map with "the scale of a mile to the mile" in Lewis Carroll's *Sylvie and Bruno Concluded*[24]. Machines also learn by compression. A central challenge of machine learning is computing low-dimensional maps that preserve the shape of high-dimensional data[25,26], though the interpretation of the dimensional reduction process is often difficult. In biological systems, learning and adaptation are linked to evolution and genetic inheritance, but also to phenotypic plasticity[7,27]. Evolution drives the emergence of compressed representation for its evident efficiency and other advantages, such as faster response and error-resilience. Smoothness and learning are two sides of the same coin: learning is only possible because the environment is smooth and therefore compressible.

**Dimensional reduction and the role of theory in biology**

Before delving into some mathematical details, we want to mention the long history of relating understanding and compression. It was recognized very early by Leibniz that [28]

> …one who acts perfectly is similar to an excellent geometer who can find the best constructions for a problem …or to a learned author who includes the greatest number of truths [réalités] in the smallest possible volume.

The quest for simple aesthetic answers in the confusing world that engulfs us—by reducing its dimensional complexity—certainly goes way back to the Greek. But it was Newton who, for us, most clearly summarized what is involved [29]:

> …[I] considered mechanics in a twofold respect: as rational, which proceeds accurately by demonstrations, and practical. To practical mechanics all the manual arts belong,… But as

> artificers do not work with perfect accuracy,… mechanics is so distinguished from geometry that what is perfectly accurate is called geometry; what is less so, is called mechanical. However, the errors are not in the art, but in the artificers…. Geometry does not teach us to draw these lines, …it is the glory of geometry that from those few principles, brought from without, it is able to produce so many things…

He makes a clear distinction between the theoretical ("geometrical") part, which is elegant, and the "engineering" (what he calls "mechanical") part, which entails a constant struggle with the details of what has to be *practically* done. All this raises the general question regarding the role of theory in biology. The program of Newton has been immensely successful in developing our understanding of the physical world. But can it be applied to living matter as well?

The problem here is two-fold. First, the biological world seems much more complex than the world of celestial mechanics. And second, currently, biological research could be classified as engineering or "mechanics," trying to understand the intricate details of living matter while avoiding the quest for universality. Perhaps the only general principle in biology is evolution: all things survive or vanish through cycles of mutation and selection. But this is a far cry from a physical law. Has the time come to apply simplifying theories to biology?— The following citation from G. Jona-Lasinio [30] summarizes well the current state of affairs:

> Theoretical physics was recognized as an independent field of research only at the end of the 19th century, shortly before the great conceptual revolutions of relativity and quantum mechanics. Today theoretical physics has multiple facets. I think that the time has come for a more precise characterization of the research field of theoretical biology, and for an assessment of its scope. [Translated from Italian]

While there is still a long way to go, ideas of dimensional reduction during the coupled evolution of genomes and phenomes seem to give some insight into how theoretical biology may progress. This essay tries to convey these general ideas by demonstrating a concrete example of protein evolution. However, to pinpoint what is special about the biological realm, we need first to describe dimensional reduction from a mathematical perspective. Once this is established, we will argue that the reduction in biology may be stronger than what general mathematical principles suggest.

## II. The curse of dimensionality and its possible cures

Often, biological data come in the form of long vectors. In sequence alignment, for example, one takes a list of $m$ genes. Each gene is a sequence of $D$ numbers taken from the alphabet of four DNA bases. A typical protein is encoded by a gene of 1000 bases in length. We can then say that we have $m$ vectors $\vec{X}_i$, $i = 1, …, m$, each representing a different sequence. Each vector has $D$ components or coordinates. The number of coordinates $D$ is the apparent or **extrinsic dimension** of the data, the dimension of the ambient space in which these $m$ vectors reside.

It is natural to ask how much information we can gain from such a measurement. An unavoidable problem when dealing with high-dimensional spaces is that any realistic sample size $m$ would be ridiculously minute compared to the size $M$ of the space, $m \ll M$. Take our sequence alignment example: Large data sets typically contain a few thousand sequences, while the combinatorial number of all possible DNA words of length $D = 1000$ is $M = 4^{1000} \approx 10^{200}$ (much more than the number photons in the cosmic background radiation, roughly $10^{90}$). Hence, one should abandon any hope to effectively sample all dimensions of this humongous space. This effect is known as:

- **The curse of dimensionality**: when the dimension of a space increases, its size explodes so fast that any data sample becomes exceedingly sparse [31-33].

But how many dimensions can we still probe with our sample? This number defines the inherent or **intrinsic dimension** of the data set, which tells us how much information it carries. To intuitively grasp what we are looking for, envision the data points as a cloud of small particles floating in a high-dimensional space of extrinsic dimension $D \gg 1$ (Figure *1*A). Each particle is situated at the position $\vec{X}_i$ corresponding to the data point. Now imagine you project the data on a low-dimensional screen of dimension $d$ by shining a flashlight. The shadow each particle casts on the screen is its projected position, $\vec{x}_i$. So we can think of the projection as a map, $\vec{X}_i \to \vec{x}_i$, from a $D$-dimensional to a $d$-dimensional space.

A projection is **reliable** if it does not deform too much the structure of the data cloud. A good measure for such a deformation is the relative change $\varepsilon$ in the distances among the points when projected on the screen. For example, if we set this tolerance to be $\varepsilon = 10\%$, then all projected distances are within $\pm 10\%$ from the original ones (Figure **1**). Thus, we can define:

- The **intrinsic dimension** of a data set is the minimal necessary dimension $d$ of the "screen" on which we can project the data reliably, i.e., with deformation smaller than a given tolerance $\varepsilon$.

### The Johnson-Lindenstrauss limit and the law of diminishing returns

Our discussion raises the natural and practical question: How does the intrinsic dimension of a data set depend on the tolerance $\varepsilon$? Intuitively, we expect that the intrinsic dimension will increase if we lower the tolerance and demand a more accurate projection, but by how much?— The seminal result by Johnson and Lindenstrauss gives an exact answer[34] (Figure *1*B):

- Any data set of $m$ points in a $D$-dimensional space can be reliably projected, with a tolerance $\varepsilon$, onto a space of only $d_* = (8/\varepsilon^2)\ln m \approx (18.4/\varepsilon^2)\log_{10} m$ dimensions. Thus, $d_*$ is the upper limit on the intrinsic dimension of any data set—no matter what its nature is.[2]

---

[2] The literature also contains many results showing that the mathematical bounds such as the $d_*$ limit are tight, i.e. they cannot be improved.

For example, assume that we have a superb data set with $m = 10^{12}$ points, then, with an error of $\varepsilon = 10\%$, these trillion points can be presented in a space of no more than $d_* \approx 2 \cdot 10^4$ dimensions. The coordinates in this projected space are linear combinations of the original ones. The intrinsic dimension, true to its name, is independent of the extrinsic dimension: $d_*$ does not depend on $D$. However, $d_*$ is quite sensitive to the tolerance $\varepsilon$, and increases like its inverse squared, $d_* \sim 1/\varepsilon^2$.

The Johnson-Lindenstrauss limit $d_*$ is extremely useful. First, it is the standard dimensional reduction we can expect to find in any data set. So if we find that the intrinsic dimension of a data set is even lower than $d_*$, we can conclude it originates from some underlying mechanism. Second, it tells us that the benefit we gain by investing more effort in increasing the data set grows very slowly due to the logarithmic dependence, $d_* \sim \log_{10} m$ (Figure *1*B). In the previous example, if one takes a million-fold smaller data set of only $m = 10^6$ points instead of $10^{12}$, then the limit only moderately decreases by 50% to $d_* \sim 10^4$. So toiling to increase the data set from a million to a trillion points might be rather futile. This implies a law of extremely **diminishing returns** on expanding the data set:

- Any doubling of the data size, $m \to 2m$, adds a constant number of dimensions to the maximal intrinsic dimension $d_*$.

For example, by doubling $m$ from $10^{12}$ to $2 \cdot 10^{12}$ we gain no more than by doubling it from 10,000 to 20,000. So the benefit from those extra ten thousand points is roughly the same as from the additional trillion points.

This widespread phenomenon of dimensional reduction of data comes under several names, depending on the authors: concentration of sets[35], the Johnson-Lindenstraus lemma[34], and the equivalence of ensembles in statistical mechanics[36-38]. While these concepts differ in their precise mathematical settings, they are essentially based on the following easy observation: Take a ball of radius $r$ in $D$ dimensions whose volume $V_D(r)$ is proportional to $r^D$. A skin of thickness $\varepsilon$ at the ball's surface has a volume $S_D(r, \varepsilon) = V_D(r) - V_D(r - \varepsilon)$. Since the ratio $V_D(r - \varepsilon)/V_D(r) = (1 - \varepsilon/r)^D$ vanishes for high dimension $D$, practically all the volume of a high-dimensional ball is at its surface, $S_D(r, \varepsilon) \approx V_D(r)$. Thus, most data points are very close to the surface of the ball, allowing for a drastic reduction in their dimension without much deformation. From all this, one should retain the observation that high-dimensional large data sets can be well approximated in low-dimensional spaces if one allows for small errors.

**The box-counting dimension**

There are many ways to estimate the intrinsic dimension of a set of points, and the books of Mattila[39] and Falconer[40] are classical references. They compare the various definitions and related methods to measure the intrinsic dimension of a set, *i.e.,* the number of variables one needs to represent it. The most popular algorithm is **box-counting**[41], which relies on the observation above regarding the skin of the hyper-sphere. One simply counts the number $N(R)$ of data points in boxes (or balls) of radius $R$. If the dimension of the set is $d$, then one observes a power law, $N(R) \propto R^d$, or a linear relation in a log-log graph, $\log N(r) \sim$ $d\log R$. For applications, there is a rule of thumb to estimate how many points $m$ are needed to measure a dimension $d$ reliably. The answer is about $m \sim 10^d$ data points[42], which can be viewed as another form of the Johnson-Lindenstrauss lemma: one cannot infer a dimension $d$ larger than the bound $d_* \sim \log_{10} m$.

In the next section, we will discuss living systems often exhibit further dimensional reduction, to an even lower dimension, $d \ll d_*$, linked to their biological function.

## III. Lowering Dimensions in Life: Smoothness and Learning

Life is high-dimensional by nature because living matter is inherently dual: all living things carry an internal image, the genome, which encodes their external aspects, their physiological traits, known as their phenome. Both genome and phenome have high dimensions: The internal image, the genome, is a long word written in a digital language with four letters, the DNA bases, and since each base can change independently, the extrinsic dimension $D$ of the genome is its length. It varies widely, ranging from a few million for bacteria to hundreds of billions in certain fish and plants. At any rate, even the million-dimensional genome of the humble bacterium is gigantic. Similarly, the corresponding spaces of phenotypes have high extrinsic dimensions. For example, the biochemical states of a microbe reside in a space whose dimension $D$ is equal to the number of chemical species, at least a few thousand.

Often, these huge dimensions are mapped to spaces of much lower dimensions. The mathematical results we discussed deal with the dimensional reduction of generic high-dimensional point sets, disregarding their specific features. But life achieves an even stronger dimensional reduction, thanks to its dual nature, which brings about two non-generic types of reduction.

The first type of reduction reflects the **smoothness** of the reality that is sensed, processed, and represented. This inherent smoothness of the physical world originates from a fundamental feature of large dynamical systems: The interactions among the many bodies that compose such systems typically give rise to large-scale collective excitations that are smooth in space and time. The resulting smoothness allows for a dimensional reduction since the relevant features of the system are captured by a small number of modes.

A simple example is a smooth 1D function that can be well-approximated by a sum of a few sinusoidal waves. The emergence of such "slow manifolds" and "soft modes"[3] is ubiquitous in dynamical systems and natural settings[43]. Think of a set of digital photographs of natural or man-made environments taken at a resolution of $1000 \times 1000$ pixels. The extrinsic dimension of the gigantic space of all possible images is the number of pixels, $D = 10^6$. However, thanks to the smoothness of physical reality, these pixel matrices are much more constrained than a random image[44,45] and can therefore be easily compressed with minimal loss of detail[46]. A more primordial example is the genetic code table, which is also smooth since nearby codons tend to encode the same or similar amino acids, thereby mitigating the impact of errors [47,48].

---

[3] Soft modes are large-scale modes whose energy is low and are therefore easy to excite.

The second type of reduction, which one may call **learning**, is through internal representation. Information about the environment and physiological state of the organism is mapped to a multi-scale internal image. These internal mappings are models of reality that distill the useful knowledge and are therefore highly compressed and low-dimensional. Such compression is common to all levels of the sensory system. However, the extent of the compression is often unclear.

One example, known since the 17$^{th}$ century, is the trichromacy of human color vision[49]: every color we sense can be matched by a mixture of three primary lights. Now we know that the three-dimensionality of color perception is reflected in the three kinds of photoreceptors in the retina, as mimicked in the design of RGB color displays of phones and computers. Another example is mammal olfactory systems that contain about 1000 types of receptors, which can potentially generate an enormous space of odor combinations[50,51]. However, psychophysics experiments reveal that smell perception space can be projected on a much lower dimension, with the key dimension being the "pleasantness" of the odor[52,53]. One may conclude from this evidence that:

- Whether we sense the electronic structure of molecules through smelling, photons through vision, or phonons through the hair cells in our ears — this high-dimensional sensory input is mapped onto a low-dimensional internal representation with a manageably small number of relevant degrees of freedom. This is performed by a hierarchical computational system combining molecular and neuronal circuitry.

Of course, these two types of reduction are intertwined and mirror each other: efficient compression by learning is feasible only thanks to the smoothness of the reality being learned. From all this, we see that, unlike generic data sets, genotypes and phenotypes cannot be just arbitrary. They are selected by evolution to be functional, leading to a further reduction of dimension relative to the bounds a general mathematical set would attain. In the protein we discuss below, smoothness shows in soft, low-energy modes that facilitate protein function. And compression by learning (adaptation) shows in the genetic representation, which appears high-dimensional but has a low-dimensional non-random component encoding the important collective modes [19-21].

## IV. An example: the low dimension of protein

Having explained the general principles of dimensional reduction in the mathematical and the biological world, we now sketch quantitatively how a model can usefully bridge the two, in the example of protein evolution[18-21]. As discussed above, the gene can be thought of as Turing tape on four DNA symbols that encodes the synthesis of a functional protein from a sequence of the 20 amino acid species in the genetic code. For our purpose, it suffices to consider a simplified version where the genes are binary sequences of 0's and 1's and to explore evolutionary dynamics of selection and mutations of these sequences.

We start by introducing the subject matter: A wide class of proteins whose functionality relies on large-scale conformational changes[54-57] (Figure 3A). These proteins typically open and close like a Pac-Man when other compounds approach and bind them[58-61]. This motion helps them catalyze or transmit long-range signals—a phenomenon called allostery[62-65]. The internal strains and stresses involved in such motions have been estimated from a comparison of bound and unbound structures[17,21].

As the forces are local, it is natural to model them within an elastic field theory, in terms of networks whose vertices are amino acids with interaction represented by springs[19,20]. In the spirit of Newton's introduction, the details should not matter, and thus one introduces simple 2D (or 3D) networks of such springs, as shown in Figure 3B. Such simplification puts the protein problem in the realm of amorphous condensed matter theory [66].

The model protein comprises two species of amino acids, polar and hydrophobic (the HP model [67]). The protein sequence is coded in a one-dimensional string of 0's (polar) and 1's (hydrophobic) — a gene $\vec{X}_i$ whose length is the number of amino acids, $D = 200$ in our example, which is also the dimension of the space of genotypes. There are $2^{200} \approx 10^{60}$ possible genes. The gene encodes the elastic network: each amino acid forms bonds (springs) with its neighbors according to a chemical interaction rule (Figure 3B).

"Evolution" randomly mutates the gene, where the functionality of the network (i.e., its "fitness") measures how well it transmits forces as a response to a local force probe mimicking binding at the active site. We ran a Monte-Carlo evolution algorithm $m = 10^6$ times, each time until a prescribed fitness level was reached. Thus, we obtained a million 200-dimensional gene vectors, $\vec{X}_1, \vec{X}_2, ....$, which encode a million different realizations of a functional protein.

The phenotype of a functional protein is captured by its low-energy modes. These motions of the protein are vectors $\vec{v}_i$ of $2 \times 200$ numbers encoding the 2D motion of each amino acid (green arrows in Figure 3B). We can now ask about how strong the dimensional reduction is when we limit our set to functional spring networks and the genes $\vec{X}_i$ representing them. As we will see, functionality entails strong restrictions on their geometry and thereby stronger concentration of the set of phenomes $\vec{v}_i$. In the set of corresponding genomes, there is a similar reduction, but only of the non-random component of the gene.

Two tools of theoretical physics prove valuable in mapping the many-body interactions among the amino acids to the evolution of the gene [21]: First, the effect of mutations on the spring network, especially on the functional mode, is conveniently formulated in terms of Green's function expansions. Second, the spectra of cross-correlations among genomes and phenomes and their comparison to those of **random matrices**. Both methods describe a huge set of samples: Either the space of all Green's functions leading to functional proteins or their representations in the space of genomes.

The **Green function** $G$ is the inverse of the elastic interaction matrix of the spring network[20]. Thus, $G$ measures the network's response to localized force impulses, which is the "fitness" of our model. The low-energy modes of the Hamiltonian describe soft functional modes, $\vec{v}_i$, such as the large-hinge-like motion of the network (Figure 3B), which mimics a similar motion of glucokinase when it binds to glucose (Figure 3A). As evolution progresses, it drives these lowest modes close to zero energy. At the same time, the mechanical response described by the Green function $G$ diverges because it costs very little energy to excite a soft mode. Therefore, we use $G$ as a direct measure of protein fitness that probes the emergence of soft functional modes. The Green function $G$ provides an explicit mapping from genotypes to phenotypes (motion vectors), $G: \vec{X}_i \rightarrow \vec{v}_i$ from which one can easily compute the fitness landscape and examine it in great detail.

**The correlation spectrum and the relevant dimension**

To extract the *relevant, non-random* component of genomes and phenomes, we use Principal Component Analysis (PCA), which measures correlations in high-dimensional spaces and gives us a first insight into the data's intrinsic structure [68]. The general idea is this: Say that we have $m$ data $\vec{X}_1, \vec{X}_2, \ldots, \vec{X}_m$, where each $\vec{X}_i$ is a $D$-dimensional vector (*e.g.,* the genes in Figure 3B). Then one calculates the correlation matrix, a $D \times D$ table of the cross-correlations among the $D$ positions in the vectors[20,21]. The $D$ non-negative real eigenvalues of the correlation matrix, $\lambda_1 \geq \lambda_2 \geq \cdots \geq \lambda_D \geq 0$, reveal the intrinsic structure of the data: the cloud of data points expands the most along directions ("principal axes") corresponding to the largest eigenvalues. Of interest here is the distribution of eigenvalues when $m$ is large, as compared to the spectrum of random matrices of the same size. The correlation spectrum $\lambda_i$ typically exhibits a bulk continuous part, describing the randomness of the data and a few outliers that capture the significant non-random correlations (Figure 4A). We will now demonstrate this revealing power of PCA for the spectra of protein evolution:

The cross-correlations matrices are computed among the set of $10^6$ genomes $\vec{X}_i$ and among the corresponding set of phenomes, the $10^6$ low modes or Green function responses $\vec{v}_i$. And we use the PCA procedure described above to examine the correlation spectra. The analysis shows clear differences between the spectral structures of coding sequences (genotypes) and functional low-modes (phenotypes).

In the case of genomes (Figure 4A top), most of the eigenvalues, $\lambda_i$, $i = 1, \ldots, D = 200$ are grouped together around an average, $\sqrt{m}/2 = 500$, while the top $d_R \approx 10$ largest eigenvalues are isolated from the bulk ("outliers"). More precisely, the bunched "continuum" spectrum spans between $(\sqrt{m} \pm \sqrt{D})/2$, and its shape is well-approximated by the Marčenko-Pastur formula for random matrices[69]. This reveals that:

- Most of the gene evolves randomly, apart from a small number of collective correlations that carry the functional information.

We see that while the intrinsic dimension of the genotype space is high $d \approx D = 200$, the dimension of the non-random component, which one may call the **relevant dimension,** is much smaller, only $d_R \approx 10 \ll d \approx D$.

The phenotypes, the low energy modes (Figure 4A bottom), have much less freedom to vary. For a mode to be soft, the motion $\vec{v}_i$ (green arrows in Figure 4B) must be as *smooth* as possible. Neighboring amino acids should move in concert. Otherwise, bonds will be strained by high shear, which will be costly in elastic energy. This severe physical constraint pushes the continuous spectrum of the phenotypes towards 0. Equivalently, one observes that random variation costs much more energy in the phenotype than in the genotype; most random genetic mutations will keep a low mode smooth. As in genotypes, there are again ~10 outliers that capture the low-dimensional space of smooth modes. In this case, $d_R \approx 10 \approx d \ll D$. One finds there is a *direct correspondence* between the non-random outliers of the genotypes and the soft mode outliers of the phenotypes (Figure 4B). Specifically, the corresponding eigenvectors of genetic correlation and shear are localized in the same amino acid regions, such that the highly deformed regions are also the strongly correlated ones [19-21].

The PCA spectra also reflect in another geometrical aspect: The $10^6$ genotype vectors $\vec{X}_i$ form a spheroid cloud in a 200-dimensional genotype space. In comparison, the $10^6$ low mode vectors $\vec{v}_i$ reside in a 400-dimensional phenotype space (because the motion of each amino acid is a 2D vector). But the phenotype "cloud" is a flattened ellipsoid with only 10 (out of 400) significant axes.

## V. Conclusions and Outlook

In summary, the protein model demonstrates the hallmarks of dimensional reduction in biology: First, the reduction is much stronger than in generic point sets. The Johnson-Lindenstrauss bound for $m = 10^6$ points, $\sim 10^4$ for $\varepsilon = 10\%$, is larger than the length of the gene $D = 200$. However, we observe a thousand-fold stronger reduction to about ten dimensions. Second, the physical origin of the reduction is the *smoothness* of the collective functional modes. Third, the reduction reflects the dual nature of living matter. While the gene remains uncompressed, its non-random component has the same dimension as the relevant phenotype space, as manifested by the equal number of non-random outliers in the correlation spectra.

Thus, the protein example demonstrates a clear geometrical picture of the genotype-to-phenotype map as a projection of the non-random information of the gene onto a low-dimensional space of smooth phenotypes (Figure 4). This picture suggests a simple recipe one may use to get a handle on the inherent structure of large sets of high-dimensional biological data ($m$ vectors in $D$-dimensional space):

1. Estimate the intrinsic dimension $d$, for example, by box counting. Compare to the expected generic reduction to the Johnson-Lindenstrauss limit $d_* = (8/\varepsilon^2)\ln(m)$. Enhanced reduction, $d \ll d_*$, indicates system-specific mechanisms.

2. Estimate the relevant dimension, $d_R$, of the non-random component of the data, by counting the outliers in the correlation spectrum. Often, $d_R \ll d$, indicating that the information-carrying component is much smaller than the overall intrinsic dimension.

3. In the case of mapping between spaces, in particular genotype-to-phenotype maps, calculate the intrinsic and relevant dimensions, $d$ and $d_R$, in both spaces. This may reveal the geometry of the information transfer by the map.

Our discussion shows that a more geometric vision of the evolution of proteins seems useful. It illustrates how the rich experience from physics and mathematics puts some biological concepts on a theoretically founded ground. We expect that the simple recipe will apply to complex biological systems whose many-body complexity is hard to analyze and understand. Examples include measurements of genetic, metabolic and signaling networks, developmental programs, immune response, and deep sequencing. In all cases, the proposed global dimensional analysis should serve as a first step to check for strong dimensional reduction and to identify the corresponding collective variables.

## Acknowledgements

This work was funded by the Institute for Basic Science, Grant IBS-R020. This publication was produced within the scope of the NCCR SwissMAP, which is funded by the Swiss National Science Foundation.

**Figures**

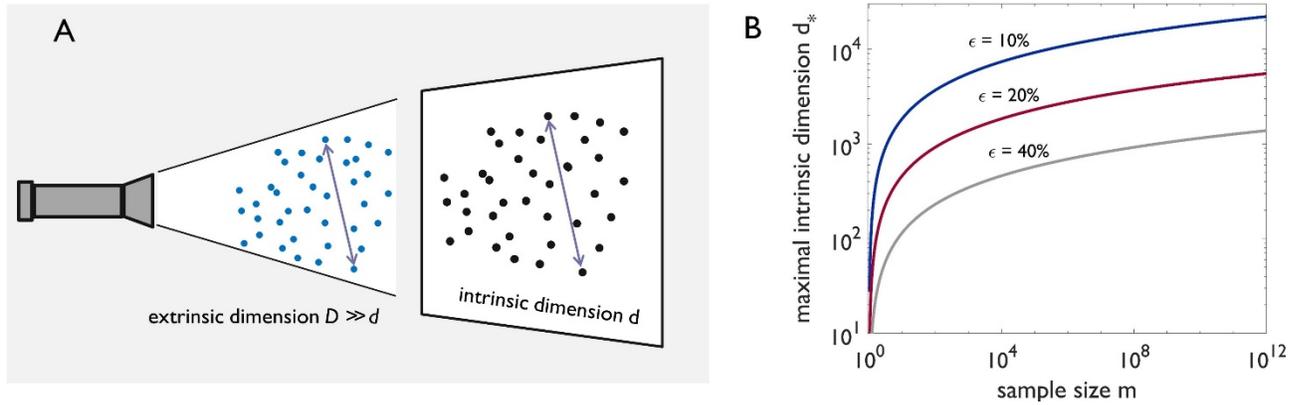

Figure **1**: **The Johnson-Lindenstrauss limit and the diminishing return on sample size.**
**A:** Data points in high-dimensional space of extrinsic dimension D are projected by a flashlight onto a low-dimensional screen of dimension $d \ll D$ The projection is reliable if it does not deform the distances (double-head arrows) by more than a set tolerance $\varepsilon$. (Note that while the point set and the flashlight are drawn on a 2D page, they are high-dimensional objects).
**B:** The Johnson-Lindenstrauss bound on the intrinsic dimension, $d_* = (18.4/\epsilon^2)\log_{10} m$, as a function of sample size m, for tolerance $\varepsilon$ = 10%, 20%, and 40 %. The slow logarithmic growth of the curves demonstrates the diminishing return on increasing the sample size.

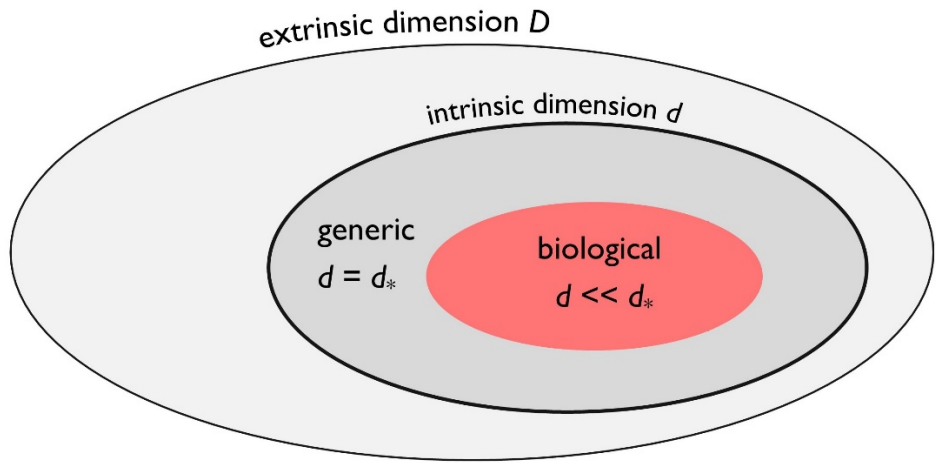

Figure 2. **Dimensional reduction in mathematics and biology**. In many living systems, the dimensional reduction from the extrinsic dimension $D$ to an intrinsic dimension d is much more drastic than the generic reduction to the Johnson-Lindenstrauss limit ($d \ll d_*$). The intrinsic dimension $d$ is the dimensionality of the space of biologically-relevant degrees of freedom that govern the evolution of the system.

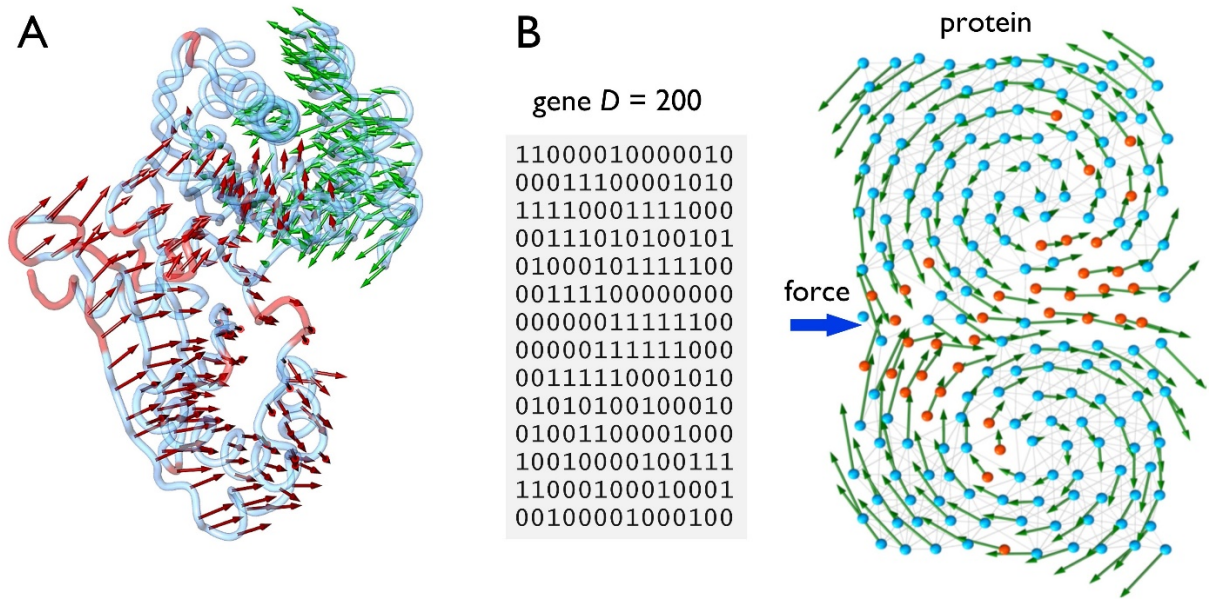

Figure 3. **Protein model and experimental motivation.**

**A:** When a glucose binds to human glucokinase, it induces low-energy, hinge-like or shear motion in the protein (arrows; calculated by comparing free and bound PDB structures, 1v4s and 1v4t). The shear deformation is high (red) in a `shear band' separating two low-shear domains (light blue) that move as rigid bodies. This topology allows large-scale conformational transitions.

**B:** The model protein (right) is made of D = *200* amino acids of two species, polar (red) and hydrophobic (blue). The protein sequence is encoded in a binary gene, a sequence $D = 200$ 0's and 1's (center). Each amino acid forms strong (gray lines) or weak (not shown) bonds with its neighbors: strong bonds between two hydrophobic amino acids and weak bonds otherwi*se* (the HP model). The fitness of the protein is the mechanical response to a localized force probe (blue arrow). After about 100 cycles of mutation and selection, the protein evolves a large-scale mechanical response: a hinge-like, low-energy motion (green arrows) with a shear band, qualitatively similar to the motion of glucokinase (A).

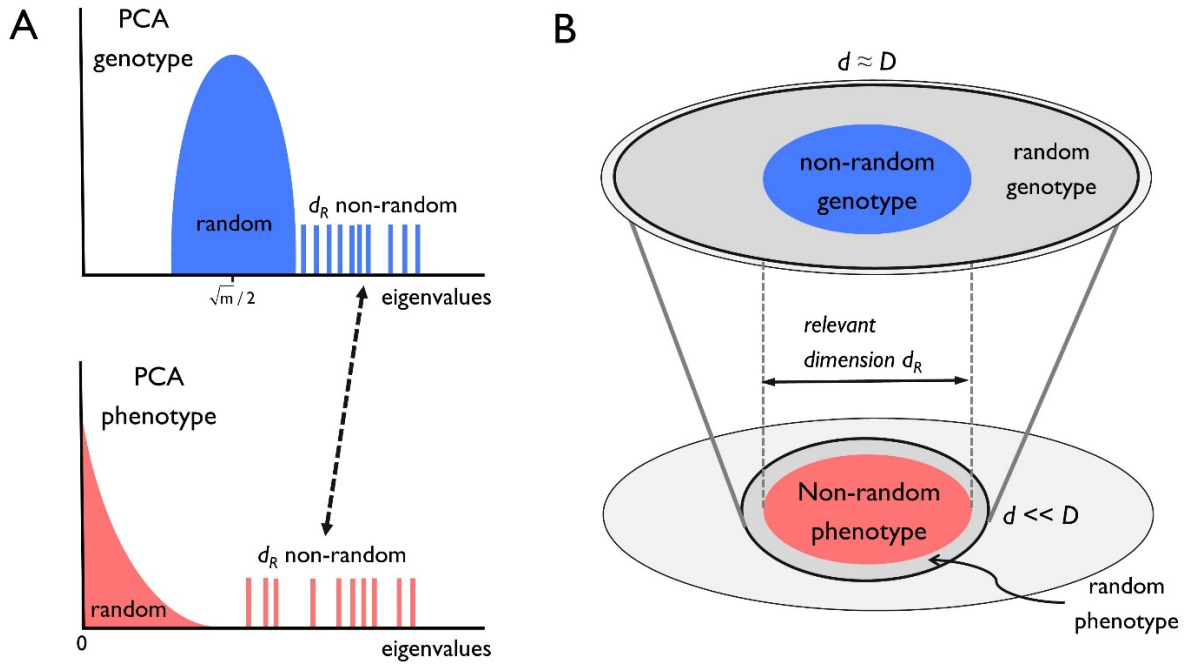

Figure 4. **PCA and the relevant dimension.**

**A:** Principal component analysis (PCA) of the correlation spectra of $m = 10^6$ genes $\vec{X}_i$ of functional proteins (blue, top), and the corresponding functional phenotypes, the large scale motions $\vec{v}_i$ (red, bottom), calculated using the Green function G (see text). Both genotypes and phenotype exhibit a bulk continuous spectrum that corresponds to the random component of the correlations. For phenotypes, this random spectrum is pushed towards 0 as the motion of the protein is much more constrained than the gene. There are $d_R \approx 10$ discrete outliers in both spectra that capture the non-random component in the evolution of protein.

**B:** A scheme of the dimensional reduction revealed by the PCA correlation spectra. The overall genotype-to-phenotype map involves huge dimensional reduction, from $d \approx D = 200$ to $d \approx 10$, due to the smoothness of the phenotype space. However, there is a one-to-one mapping between the non-random components of genes and motions, both of relevant dimension $d_R \approx 10$, reflecting the dual nature of biological systems.